\documentclass[prb,twocolumn,superscriptaddress]{revtex4}

\usepackage{graphicx}

\newcommand\rhom{\rho_\mathrm{mag}}

\begin{document}

\title{First-principles analysis of spin-disorder resistivity of Fe and Ni}

\author{A. L. Wysocki}
\email{awysocki@bigred.unl.edu}
\affiliation{Department of Physics and Astronomy
and Nebraska Center for Materials and Nanoscience, University of
Nebraska--Lincoln, Lincoln, Nebraska 68588, USA}
\author{R. F. Sabirianov}
\affiliation{Department of Physics and Nebraska Center for Materials and
Nanoscience, University of Nebraska-Omaha, Omaha, Nebraska 68182, USA}
\author{M. van Schilfgaarde}
\affiliation{Department of Chemical and Materials Engineering, Arizona State
University, Tempe, Arizona 85287, USA}
\author{K. D. Belashchenko}
\affiliation{Department of Physics and Astronomy and Nebraska Center for
Materials and Nanoscience, University of Nebraska--Lincoln, Lincoln, Nebraska
68588, USA}

\date{\today}

\begin{abstract}
Spin-disorder resistivity of Fe and Ni and its temperature dependence are
analyzed using noncollinear density functional calculations within the
supercell method. Different models of thermal spin disorder are considered,
including the mean-field approximation and the nearest-neighbor Heisenberg
model. Spin-disorder resistivity is found to depend weakly on magnetic
short-range order. If the local moments are kept frozen at their
zero-temperature values, very good agreement with experiment is obtained for
Fe, but for Ni the resistivity at elevated temperatures is significantly
overestimated. Agreement with experiment for Fe is improved if the local
moments are iterated to self-consistency. The overestimation of the resistivity
for paramagnetic Ni is attributed to the reduction of the local moments down to
$0.35 \mu_B$. Overall, the results suggest that low-energy spin fluctuations in
Fe and Ni are better viewed as classical rotations of local moments rather than
quantized spin fluctuations that would require an $(S+1)/S$ correction.
\end{abstract}

\maketitle

\section{Introduction}

Electron scattering off of spin fluctuations in magnetic metals
results in an ``anomalous'' contribution to electric
resistivity.\cite{Coles,Mott,Vonsovskii} The analysis of this
spin-disorder resistivity (SDR) is of interest because it can
provide material-specific information on the character of spin
fluctuations which is not easily accessible by other means.
Scattering on spin disorder is also an important factor degrading
the performance of magnetoresistive nanostructures in spintronic
devices.

The magnitude of the spin-disorder contribution to resistivity is
comparable to the phonon contribution near and above the Curie
temperature $T_c$.\cite{Coles} (Magnetic scattering amplitudes have
no small parameter unless the local moments are small.) It is
usually assumed that SDR is constant well above $T_c$. In this
region Matthiessen's rule is valid, and the phonon contribution can
be fitted to the Bloch-Gr\"uneisen formula. The excess resistivity
in the whole temperature range may be attributed to spin
disorder,\cite{Weiss} although one may expect deviations from
Matthiessen's rule at low temperatures where transport is carried by
weakly interacting spin channels.\cite{Fert} In addition, it was
argued that in some cases (such as Ni) spin disorder may change the
character of states on the Fermi level and thereby appreciably
change the phonon contribution itself.\cite{Coles,Mott}

Many authors have studied SDR theoretically using the $s$-$d$ model
Hamiltonian.\cite{Kasuya,dGF,Mannari,VI} In this approach the $3d$ shells in
transition metals (or $f$ shells in rare earth metallic magnets) are assumed to
be localized at atomic sites and partially filled, forming magnetic moments
$\hat\mathbf{S}_i$ that are coupled to the current-carrying conduction
electrons by exchange interaction
$H_{sd}=-J_{sd}\sum_i\hat\mathbf{S}_i\hat\mathbf{s}_i$, where $J_{sd}$ is the
local $s$-$d$ exchange coupling constant and $\hat\mathbf{s}_i$ the
spin-density operator of the conduction electrons at site $i$. Thermal
fluctuations of the $d$-electron spins generate an inhomogeneous exchange
potential; in the Born approximation the SDR is then determined by the
conduction electron band structure and the spin-spin correlation functions of
$d$-electron spins.\cite{VI} If the scattering is approximated as being
elastic, only equal-time spin correlators have to be considered. Further, if
the mean-field approximation (MFA) is used for $3d$ spin statistics, the SDR
behaves as $\rhom(T)=\rho_0[1-M^2(T)/S(S+1)]$, where
$M(T)=\langle\mathbf{S}(T)\rangle$ is the magnetization at temperature $T$ and
$\rho_0\propto J_{sd}^2S(S+1)$. \cite{Kasuya} Note that above $T_c$ SDR is
constant and equal to $\rho_0$. The shape of the Fermi surface of conduction
electrons is immaterial to this prediction as long as the scattering is
elastic. \cite{VI}

The effects of magnetic short-range order (MSRO) on SDR have also been
investigated within the $s$-$d$ model.
\cite{dGF,Mannari,Fisher,Gibson,Rossiter,Alexander,Kataoka,Akabli} This problem
has attracted considerable attention in connection with a ``bump'' in the
resistivity that is observed near $T_c$ in some magnetic metals (although it is
usually quite small).\cite{Coles} The analysis of critical MSRO effects showed
that a cusp may appear near $T_c$ due to long-wave critical
fluctuations,\cite{dGF} although it should usually be strongly suppressed by
finite mean-free path and cancelations due to Fermi surface
integration.\cite{Fisher} It was also found that the effect of MSRO and even
its sign are sensitive to such details of the model as the conduction band
occupation and the form of the scattering
(pseudo)potential.\cite{Rossiter,Alexander,Kataoka}

Although the $s$-$d$ model provided physical insight into the mechanism of SDR,
it suffers from serious limitations. First, the distinction between localized
and conduction electrons is not justified in transition metals where $3d$
electrons are itinerant and form the Fermi surface. Even if the current is
dominated by light $s$-like bands that can be distinguished from heavy $d$-like
bands, the relaxation rate is dominated by interband ($s$-$d$) scattering.
\cite{Goodings} Second, at elevated temperatures the scattering potential
generated by spin disorder is of the order of the exchange splitting, which is
not small compared to the bandwidth. This invalidates the Born approximation
which is usually made in model calculations. Third, the $s$-$d$ model does not
properly take into account the change of electronic structure due to disorder.

The first-principles approach to SDR is free from all these limitations and can
be used for quantitative calculations of SDR. This is of particular interest
for the theory of itinerant magnets, because, as mentioned above, SDR depends
on spin-spin correlation functions. Different theories of itinerant magnetism
make conflicting predictions for such properties as the degree of MSRO, the
mean-squared magnetic moment, and their temperature
dependence;\cite{Moriya,Wang,Antropov,Ruban,WGB} these quantities are quite
hard to measure directly. By calculating SDR for a particular model of spin
fluctuations and comparing the results with experiment, one can attempt to
validate or rule out different spin-fluctuation models.

Earlier we have calculated the temperature dependence of SDR in Fe and Ni using
supercell calculations within the tight-binding linear muffin-tin orbital
(TB-LMTO) method using the mean-field distribution for spin orientation
statistics. \cite{Wysocki} Good agreement with experiment was obtained for Fe,
but for paramagnetic Ni the SDR was found to be significantly overestimated. In
this paper we analyze the temperature dependence of SDR for Fe and Ni in
greater detail and consider the effects of magnetic ordering, MSRO, and local
moment reduction. We also study the importance of the basis set size and
self-consistency of the atomic potentials.

\section{General approach and methods}\label{general}

Our approach is based on noncollinear density functional theory (DFT). All the
valence electrons are treated on the same footing, while the scattering
potentials are determined by the self-consistent electron charge and spin
densities. We use the TB-LMTO method \cite{Andersen} which represents the
electronic density of the crystal as a superposition of overlapping atomic
spheres; the electronic density inside each sphere is spherically symmetric.
This method is known to work very well in close-packed materials, and it allows
us to introduce spin disorder in various ways. In this work we used the rigid
spin approximation which assumes that the spin density in each atomic sphere
remains collinear, while the spin densities of different atomic spheres become
noncollinear at finite temperatures. In the simplest model the electron charge
and spin densities in all atomic spheres are taken from the ground state and
frozen, while the directions of the spin moments in different spheres are
randomized with the angular distribution function taken from MFA at the given
temperature. This model is expected to work reasonably well for Fe which has a
fairly stable local moment.\cite{Gyorffy} In Section \ref{MFA} we show that
this is indeed the case; however, for Ni the paramagnetic SDR calculated in
this way is about twice too large. In order to explain this discrepancy, the
dependence of SDR on the degree of MSRO and on the magnitude of the local
moment is studied in Sections \ref{MSRO} and \ref{reduc}.

\begin{figure}[htb]
\includegraphics*[width=0.45\textwidth]{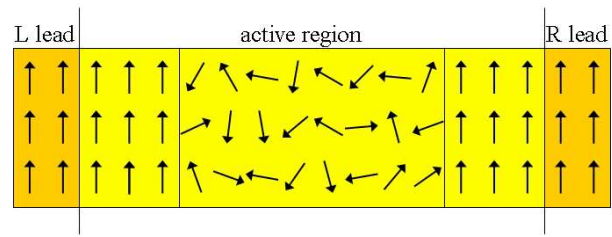}
\caption{(Color online) The schematic picture of the system used in the
calculations. Vertical lines indicate the embedding planes.}\label{fig1}
\end{figure}

We use the supercell approach and calculate the areal conductance of a layer of
spin-disordered metal FM(D) sandwiched between fully ordered semi-infinite
leads FM(O) made of the same metal (see Fig.\ \ref{fig1}). The resistivity is
then proportional to the slope of the inverse conductance as a function of the
disordered layer thickness, once the Ohmic limit is reached. For the given
thickness of the FM(D) layer, the conductance of the system was averaged over
several disorder configurations (typically 15). The planar system is
represented by a laterally periodic prism with an axis along the [001]
crystallographic direction, and care is taken to make sure that the conductance
scales as the cross-section of the prism. To calculate the conductance we use
the principal-layer Green's function technique \cite{Turek,Kudrnovsky} and the
Landauer-B\"uttiker formalism \cite{Datta} in the implementation allowing for
noncollinearity in the active region.\cite{MvS} This technique was employed
before to study the effects of substitutional disorder on transport in magnetic
multilayers; \cite{Drchal} it is similar to the supercell Kubo-Greenwood method
used to calculate the residual resistivity of binary alloys.\cite{Brown}

If the atomic potentials in the supercell are not converged to self-consistency
with the given spin disorder configuration, care needs to be taken to ensure
local charge neutrality. Indeed, FM(D) and FM(O) materials have different Fermi
levels that must normally be matched by the contact voltage. In order to
enforce charge neutrality in the FM(D) region, a constant potential shift was
introduced in this region so that the charge in the central part of FM(D)
averaged over disorder realizations was zero. This potential shift plays the
role of the contact voltage. Note that no matter how the FM(O)/FM(D) interfaces
are treated (self-consistently or not), they add contact resistances to the
circuit. However, since the \emph{resistivity} of the FM(D) material is
extracted from the thickness dependence of the resistance in the Ohmic limit,
the simplified treatment of interfaces has no effect on the results.

\section{Spin-disorder resistivity in the mean-field approximation}\label{MFA}

\subsection{Paramagnetic state}

In this section we analyze the temperature dependence of SDR for iron and
nickel using MFA for thermal spin disorder; the spin-spin correlator is
purely local in this approximation. First we consider the paramagnetic state
where the angular distribution function is isotropic, and the resulting SDR is
temperature-independent.

We need to make a physically reasonable choice of atomic potentials
for the conductance calculations. It is known that the local moments
in Fe are quite stable;\cite{Moriya} in particular, the DLM method,
which employs the coherent potential approximation for
spin-disordered states, shows only a small reduction of the local
moment in paramagnetic Fe compared to its ground-state
value.\cite{Gyorffy} As seen below, direct averaging of
self-consistent local moments in the paramagnetic states gives a
similar result. Therefore, for Fe it is reasonable to use frozen
atomic potentials taken from the zero-temperature ground state in
all calculations. We have also checked the effect of
self-consistency on SDR in Fe and found it to be small (see below).
The situation is entirely different for Ni, where the local moment
depends on the magnetic state; in particular, it vanishes altogether
in the paramagnetic DLM approximation. \cite{Staunton} Since
longitudinal spin fluctuations (that are absent in our approach) can
at least partially restore the local moments, \cite{Moriya} it is
not \emph{a priori} clear how the atomic potentials should be
modified for Ni. In this section we use frozen atomic potentials;
the necessary corrections are discussed later.

\begin{figure}[htb]
\includegraphics*[width=0.45\textwidth]{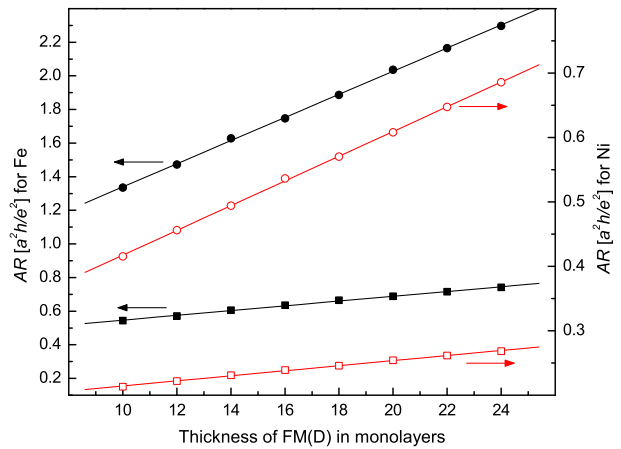}
\caption{(Color online) The area-resistance product $AR$ of the
FM(O)/FM(D)/FM(O) systems as a function of the FM(D) layer thickness for bcc Fe
(black filled symbols) and fcc Ni (red or gray empty symbols) obtained with
$l_{max}=2$. Circles and squares correspond, respectively, to the paramagnetic
state and to the lowest temperature for which the calculations were made.
$4\times4$ and $3\times3$ supercells were used for Fe and Ni, respectively,
with edges along the [100] directions. Straight lines show the linear fitting;
error bars are smaller than the size of the symbols.} \label{fig2}
\end{figure}

Fig.\ \ref{fig2} shows the inverse areal conductance for paramagnetic Fe and Ni
as a function of the disordered FM(D) region thickness. Here we used the frozen
ground-state atomic potentials and the LMTO basis including $s$, $p$, and $d$
orbitals ($l_{max}=2$). The supercell cross-sections contained $4\times4$ (for
Fe) and $3\times3$ (for Ni) cubic unit cells with edges oriented along the
[100] directions. Almost perfect Ohmic behavior is apparent for both Fe and Ni,
which establishes the validity of our approach.

Table ~\ref{tab1} lists the values of SDR found for paramagnetic Fe and Ni
using different supercell cross-sections, LMTO bases truncated at $l_{max}=2$
and $l_{max}=3$ (the latter includes $f$ orbitals), as well as the value found
using self-consistent (rather than frozen) atomic potentials for Fe. It is
seen that the results are well converged with respect to the supercell
cross-section, and even $2\times2$ supercells provide sufficient accuracy. This
is reasonable because the mean-free path in the paramagnetic state is small.

\begin{table}
\caption{Spin-disorder resistivity in $\mu\Omega\cdot$cm for paramagnetic bcc
Fe and fcc Ni. The calculated values are given for basis sets with $l_{max}=2$
and 3, as well as for different lateral cell sizes with edges along the [100]
directions. SC denotes calculations with self-consistent potentials. Standard
deviations of SDR due to limited disorder sampling are included.} \label{tab1}
\begin{ruledtabular}
\begin{tabular}{lccccc}
Metal and basis & $M$, $\mu_B$ & $2\times2$ & $3\times3$ & $4\times4$ & Exp.\cite{Weiss}\\
\hline
\hbox{Fe:} $l_{max}=2$ & 2.29 & $106\pm1.8$ & $101\pm1.3$ & $102\pm1.0$& $80$\\
\phantom{Fe:} $l_{max}=3$ & 2.22 & $86\pm1.6$ & $87\pm7.1$ & $85\pm7.4$ & $80$\\
\phantom{Fe:} $l_{max}=2$, SC & 2.21 & $88\pm3.7$ & & & $80$\\
\hline
\hbox{Ni:} $l_{max}=2$  & 0.66 & $34\pm0.6$ & $35\pm0.4$ & & $15$\\
\hline
\phantom{Ni:} $l_{max}=3$ & 0.63 & $29\pm0.6$ & & & $15$\\
\end{tabular}
\end{ruledtabular}
\end{table}

\begin{figure}[htb]
\begin{center}
\includegraphics[width=0.45\textwidth]{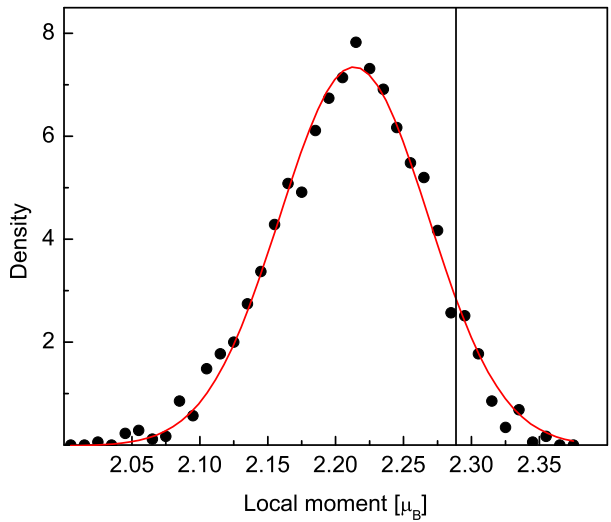}
\end{center} \caption{(Color online) Distribution of the local magnetic moment in self-consistent fully
spin-disordered bcc Fe. The Fermi temperature is equal to the experimental
$T_c$. The vertical line shows the local moment at $T=0$. The red (solid) curve
shows the Gaussian fit to the data.} \label{Fe-moment}
\end{figure}

The calculations with self-consistent atomic potentials were performed as
follows. In order to reduce the statistical error, the averaging of the
conductance was performed using the same sets of random spin disorder
configurations as in the calculation with frozen potentials. For each
individual spin configuration the atomic potentials were iterated to
self-consistency using the Fermi distribution function corresponding to the
experimental $T_c$ of Fe. The resulting distribution of the sites over the
magnitude of the local magnetic moment is shown in Fig.\ \ref{Fe-moment}; this
distribution is Gaussian with a rather small width. The average local moment is
only reduced by 3-4\% from its ground state value. This small reduction appears
to be similar to the DLM calculations of Ref.\ \onlinecite{Ruban}, while Ref.\
\onlinecite{Gyorffy} obtained a somewhat larger reduction. The self-consistent
density of states (not shown) is very similar to the one generated by the
frozen ground-state atomic potentials (see Fig.\ \ref{Fe-dos}e below).

The addition of $f$ orbitals to the LMTO basis reduces the
calculated SDR by approximately 15\% for both Fe and Ni.
Self-consistency in the paramagnetic state of Fe results in a
similar reduction. This similarity suggests that the main reason for
this SDR decrease is the reduction of the local moment, which is,
incidentally, very similar in both cases. In order to check this, we
performed additional calculations for Fe in which the $f$ channel
was added to the basis while the charge density was kept unchanged
from the self-consistent one with $l_{max}=2$. For the frozen
potential case, SDR reduced slightly from 106 to 100
$\mu\Omega\cdot$cm; for the self-consistent paramagnetic case, it
only reduced from 88 to 86 $\mu\Omega\cdot$cm, which is within the
error bar. Thus, the effect of $l_{max}$ \emph{per se} on SDR is
very small for Fe. This is somewhat different from the binary alloy
systems considered by other authors using both TB-LMTO and KKR
(Korringa-Kohn-Rostocker) methods, where a larger effect of adding
$f$ states was found. \cite{Turek2,Banhart} In view of the weak
dependence of SDR on $l_{max}$, below we use $l_{max}=2$ in all
calculations for $T<T_c$.

The experimental estimates of SDR in the paramagnetic state \cite{Weiss} are
listed in the last column of Table \ref{tab1}. The agreement with experiment
for Fe is quite satisfactory, and it is in fact improved if the reduction of
the local moment is included. In Ni the SDR calculated with frozen atomic
potentials is overestimated by a factor of 2. This is not surprising, because,
as mentioned above, the use of frozen atomic potentials is not justified for
Ni. In order to understand the origins of the disagreement with experiment for
Ni, possible modifications of the statistical model for the paramagnetic state
must be considered; this is done below in Sections IV and V.

Recently, Buruzs \emph{et al.} \cite{Buruzs} calculated the SDR for Fe and Co
using the disordered local moment (DLM) approach within the
Korringa-Kohn-Rostocker method and found that their method significantly
overestimates the paramagnetic SDR in these metals. The source of disagreement
with our supercell method for Fe is unknown to us.

\subsection{Temperature dependence in the ferromagnetic state}

In this section we consider ferromagnetic state of Fe and Ni. We use frozen
ground-state potentials and the basis with $l_{max}=2$. As mentioned above,
this approximation is reasonable for Fe, while for Ni it is not applicable at
high temperatures; nevertheless, comparison of these two systems will allow us
to draw important conclusions. For the ferromagnetic state the spin
configurations were generated using the mean-field distribution function

\newcommand{\bmu}{\mbox{\boldmath$\mu$}}
\begin{equation}
p(\theta) \propto e^{-\mathbf{H}_{\mathrm{eff}}\cdot \bmu /T}\,,\quad
H_{\mathrm{eff}}(T)=\frac{3M(T)T_c}{\mu M(0)} \label{distrib}
\end{equation}
where $\theta$ is the angle between the local moment $\bmu$ and the
magnetization axis, $M(T)$ is the magnetization at temperature $T$ in MFA, and
$\mathbf{H}_{\mathrm{eff}}$ is the Weiss field. This distribution function
depends only on $T/T_c$.

\begin{figure}[htb]
\begin{center}
\includegraphics[width=0.35\textwidth]{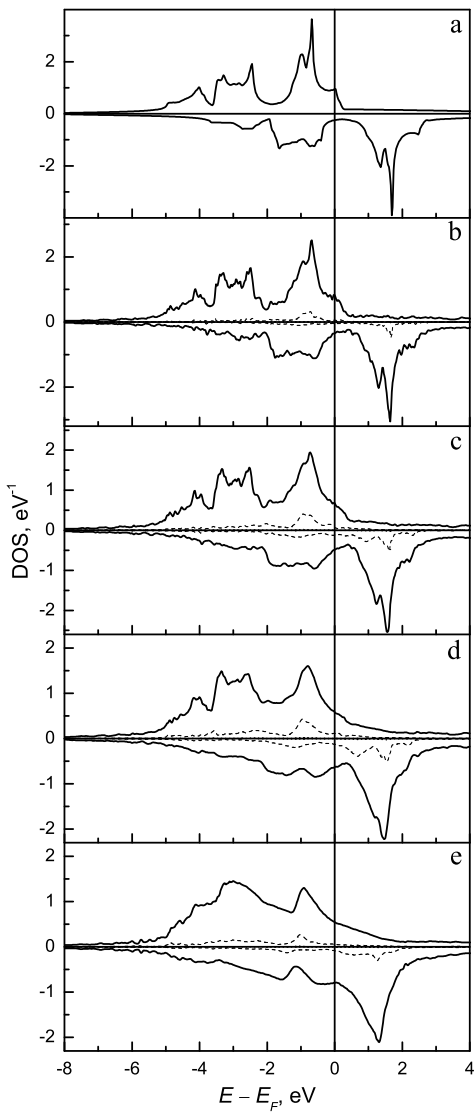}
\end{center}
\caption{Spin-resolved density of states (solid lines) for bcc Fe averaged over
random spin configurations with the mean-field distribution function
(\ref{distrib}); (a) $T=0$, (b) $T=0.25T_c$, (c) $T=0.5T_c$, (d) $T=0.75T_c$,
and (e) $T=T_c$. Dashed lines show the mean-square deviation of the DOS on a
given site from its ensemble average.} \label{Fe-dos}
\end{figure}

\begin{figure}[htb]
\begin{center}
\includegraphics[width=0.35\textwidth]{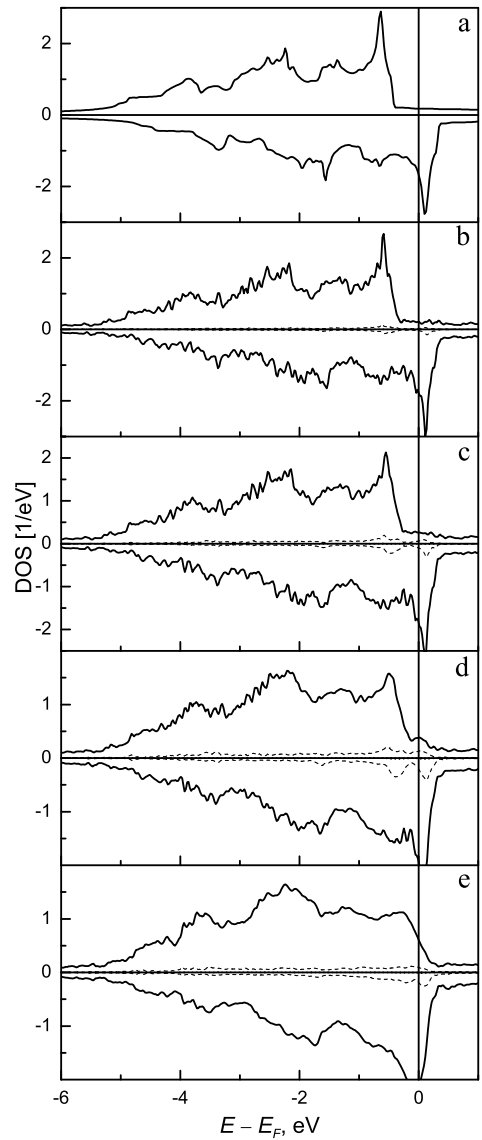}
\end{center}
\caption{Same as in Fig.\ \ref{Fe-dos} but for fcc Ni.} \label{Ni-dos}
\end{figure}

Before we turn to the temperature dependence of SDR, let us look at the
electronic structure of Fe and Ni with spin disorder. The spin-resolved DOS of
Fe and Ni is shown in Figs.\ \ref{Fe-dos} and \ref{Ni-dos} for several
temperatures. These data were obtained by projecting the site-resolved DOS onto
local spin-up and spin-down states (in the local reference frame where the $z$
axis is parallel to the local moment) and subsequent averaging over bulk-like
sites and spin disorder configurations generated according to Eq.\
(\ref{distrib}). The paramagnetic DOS of Fe is very similar to the KKR-DLM
results. \cite{Gyorffy} As the temperature is increased from 0 to $T_c$, the
spin-up and spin-down states randomly hybridize with each other, the peaks
broaden, and the van Hove singularities are washed out. The mean-squared
deviation of the DOS from its average (shown by dashed lines) is quite small,
which is a direct consequence of the large coordination number. In Fe the spin
splitting is almost independent on temperature, while in Ni it is much reduced
as $T$ gets close to $T_c$. Note that the frozen atomic potentials in Ni are
very far from self-consistency at elevated temperatures, but a self-consistent
treatment neglecting longitudinal spin fluctuations would be meaningless. We
will return to this issue in Section V.

Let us now discuss the temperature dependence of SDR. While we found above that
$2\times2$ supercells were sufficiently large for the paramagnetic state,
additional care needs to be taken at lower temperatures where the mean-free
path becomes longer. We found that $4\times4$ supercells for Fe and $3\times3$
for Ni demonstrate linear dependence of the conductance on the length of the
supercell for all temperatures down to about $T_c/3$ (see Fig.\ \ref{fig2}).
This behavior agrees with a simple mean-free path estimate using the
free-electron formula $l = \frac{3}{4}AR_{\mathrm{bal}}/\rho$, where
$AR_{\mathrm{bal}}$ is the ballistic area-resistance product; $l$ does not
exceed the lateral cell size in this temperature range. Another indication of
the Ohmic behavior comes from the distribution of the current over the spin
channels. The conductance of the FM(O)/FM(D)/FM(O) system is a sum of four
partial conductances, $G_{\uparrow\uparrow}$, $G_{\downarrow\downarrow}$,
$G_{\uparrow\downarrow}$, $G_{\downarrow\uparrow}$ (the latter two are equal).
Spin-conserving and spin-flip scattering have similar rates in our
spin-disorder problem (as long as the temperature is not too low), and
therefore the electrons ``forget'' their spin over their mean-free path.
Therefore, in the Ohmic limit the partial conductances must be proportional to
the number of channels in the left and right leads for the corresponding spin
channels: $G_{\sigma\sigma^\prime}\propto M^L_\sigma M^R_{\sigma^\prime}$. This
implies that in this regime we should have
$G_{\uparrow\uparrow}G_{\downarrow\downarrow}=G_{\uparrow\downarrow}G_{\downarrow\uparrow}$.
This relation does indeed hold down to $T\sim T_c/3$ unless the thicknesses of
the FM(D) region is very small.

\begin{figure}[htb]
\begin{center}
\includegraphics[width=0.45\textwidth]{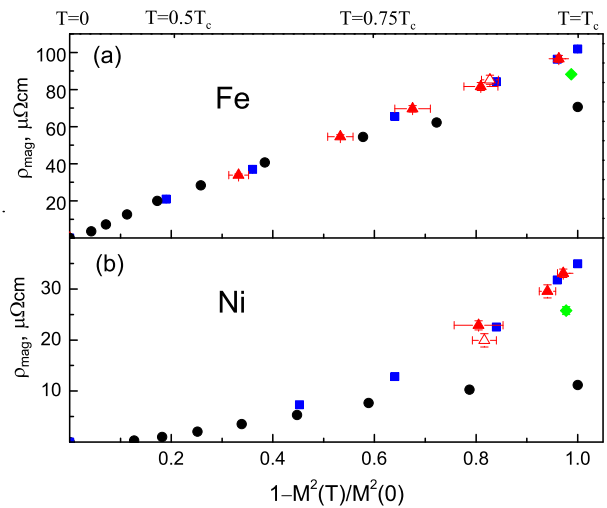}
\end{center}
\caption{(Color online) Dependence of spin-disorder resistivity on the
magnetization for (a) Fe, and (b) Ni. Black circles denote experimental data
combining Ref.\ \onlinecite{Weiss} for $\rho_\mathrm{mag}(T)$ and Ref.\
\onlinecite{Crangle} for $M(T)$. Blue (black) squares show mean-field
calculations, filled red (gray) triangles denote Monte Carlo results, and green
(gray) diamonds show reverse Monte Carlo calculations. The empty red (gray)
triangles show Monte Carlo results with larger cells: $6\times6$ for Fe and
$4\times4$ for Ni. The upper axis shows temperatures corresponding to the given
magnetization in MFA. The error bars along the $x$ and $y$ axes show
statistical uncertainties where they exceed the size of the symbols. All
results are for $l_{max}=2$.} \label{SDR}
\end{figure}

The dependence of the calculated SDR for Fe and Ni on the magnetization is
plotted in Fig.\ \ref{SDR} along with the experimental data \cite{Weiss} (those
for $M(T)$ were taken from Ref. \onlinecite{Crangle}). The results for Fe agree
rather well with experiment, especially at lower temperatures where the
magnetic excitations are dominated by spin waves and our classical approach is,
strictly speaking, invalid. This surprising finding is due to the fact that SDR
in Fig.\ \ref{SDR} is plotted as a function of the long-range order parameter
and that, as we show below in Section \ref{MSRO}, the SDR in Fe is insensitive
to MSRO. The calculated SDR exhibits linear dependence on $M^2(T)$ up to $T_c$,
while the experimental data deviate downward from the straight line. This
deviation may be attributed to a small reduction of the local moment at
elevated temperatures, as discussed in the previous section.

For Ni the deep low-temperature region could not be accessed due to the
increased mean-free path. Still, the agreement with experiment at lower
temperatures is good, while at higher temperatures the calculated SDR strongly
deviates upwards from experimental data. This deviation indicates the
inadequacy of our spin fluctuation model; its possible modifications are
studied in the following sections.

The qualitative features of the \emph{calculated} temperature dependence of SDR
(with frozen atomic potentials) are different for Fe and Ni. It is seen in
Fig.\ \ref{SDR} that for Fe the SDR is proportional to $1-M^2(T)/M^2(0)$ in
agreement with the predictions of the $s$-$d$ model if spin fluctuations are
treated classically. On the other hand, for Ni this relation does not hold. As
mentioned in the Introduction, the change of electronic structure due to spin
disorder may lead to deviations from $s$-$d$ model predictions.

As seen in Figs.\ \ref{Fe-dos} and \ref{Ni-dos}, the densities of states change
quite appreciably with temperature for both Fe and Ni. Therefore, it may seem
surprising that for Fe the temperature dependence of SDR agrees with the
$s$-$d$ model. Still, one can understand the difference between Fe and Ni using
the following considerations. First, as seen in Fig.\ \ref{Ni-dos}, the
exchange splitting in Ni is strongly reduced at elevated temperatures, which
results in the lifting of the heavy majority-spin $3d$ bands up to the Fermi
level. Scattering into these final states from the light bands becomes
possible, which decreases the lifetime of the latter. This mechanism was
invoked by Mott \cite{Mott} to argue that the reduction of the spin splitting
in Ni can result in an anomalous temperature dependence of the \emph{phonon}
resistivity. The same argument applies to SDR considered here. According to
Fig.\ \ref{Ni-dos}, this happens approximately at $T=0.75T_c$, which roughly
corresponds to the upturn of SDR seen in Fig.\ \ref{SDR}b. On the other hand,
for Fe, as seen in Fig.\ \ref{Fe-dos}, the exchange splitting is constant and
no new bands appear at the Fermi level. Consequently, no additional temperature
dependence is introduced and SDR scales as $1-M^2(T)/M^2(0)$.

While plausible, the above arguments are not conclusive, because they assume
without proof that the scattering matrix elements between the light and heavy
bands are large. On a more subtle level, one may speculate that the difference
between Fe and Ni can be understood based on the relation between disorder
broadening and spin splitting. At the given wavevector, the spectral function
consists of delta-function peaks corresponding to majority and minority-spin
states. In the presence of spin disorder, the spin states on neighboring sites
are allowed to hybridize with random matrix elements, and the delta-function
peaks broaden. At low temperature the broadening is small, and the peaks
corresponding to different spins are well separated in energy from each other.
However, at higher temperatures some of these peaks can merge and form common,
``virtual-crystal-like'' bands. Calculations of the paramagnetic spectral
functions using the DLM method indicate that in Fe the majority and
minority-spin states remain separated through large portions of the Fermi
surface even above $T_c$.\cite{Staunton85} On the other hand, in Ni the
majority and minority-spin states are mixed in the paramagnetic
state.\cite{Staunton85} Therefore, at certain temperature below $T_c$ there is
a crossover from separated to mixed-spin bands. The lifetime is expected to
decrease as the bands merge, which again explains the upturn of SDR from the
straight line in Fig.\ \ref{SDR}b.

\section{Effect of magnetic short-range order}\label{MSRO}

\begin{table*}
\caption{Spin-spin correlators $C_{0i}$ for the first three shells of nearest
neighbors ($i=1,2,3$) and the local correlator ($i=0$) are shown for Fe and Ni
for each considered temperature in Monte Carlo simulations and for reverse
Monte Carlo method. The values of the correlators were found using $4\times4$
cells and averaging over the lengths that were used in transport calculations.
The Curie temperature for bcc and fcc nearest-neighbor Heisenberg model was
obtained using the fourth order cumulant method. The values of SDR are compared
with MFA results corresponding to the same $M^2$. The listed uncertainties are
due to the limited disorder sampling.} \label{tab2}
\begin{tabular}{|c|c|c|c|c|c|c|c|}
\hline
Metal, & $T/T_c$ & \multicolumn{4}{c|}{$C_{0i}=\langle \mathbf{e}_0\mathbf{e}_i\rangle-\langle \mathbf{e}_0\rangle\langle \mathbf{e}_i\rangle$} & \multicolumn{2}{c|}{$\rho_\mathrm{mag}$, $\mu\Omega\cdot$cm} \\
\cline{3-8}
cross-section & & $i=0$ & $i=1$ & $i=2$ & $i=3$ & MC or RMC & MFA \\
\hline \hline
Fe, $4\times4$ & $\infty$ & $1$ & $0$ & $0$ & $0$ & $101.9\pm1.0$ & $101.9\pm1.0$ \\
\hline
Fe, $4\times4$ & $1.22$ & $0.96\pm0.01$ & $0.15\pm0.01$ & $0.07\pm0.01$ & $0.04\pm0.01$ & $96.6\pm1.9$ & $97.6\pm0.5$ \\
\hline
Fe, $4\times4$ & $0.98$ & $0.81\pm0.03$ & $0.14\pm0.03$ & $0.08\pm0.03$ & $0.04\pm0.03$ & $81.6\pm2.1$ & $82.4\pm0.4$ \\
\hline
Fe, $6\times6$ & $0.98$ & $0.83\pm0.02$ & $0.16\pm0.02$ & $0.09\pm0.02$ & $0.06\pm0.02$ & $81.4\pm2.3$ & $84.4\pm0.4$ \\
\hline
Fe, $4\times4$ & $0.85$ & $0.68\pm0.04$ & $0.14\pm0.04$ & $0.09\pm0.04$ & $0.07\pm0.04$ & $69.6\pm1.8$ & $69.1\pm0.4$ \\
\hline
Fe, $4\times4$ & $0.73$ & $0.53\pm0.03$ & $0.09\pm0.03$ & $0.05\pm0.03$ & $0.03\pm0.03$ & $54.6\pm1.8$ & $53.9\pm0.3$ \\
\hline
Fe, $4\times4$ & $0.49$ & $0.33\pm0.02$ & $0.06\pm0.02$ & $0.04\pm0.02$ & $0.02\pm0.02$ & $33.8\pm0.7$ & $33.6\pm0.2$ \\
\hline
Fe, $4\times4$ & RMC & $0.99\pm0.01$ & $0.30\pm0.01$ & $0.16\pm0.01$ & $0.05\pm0.01$ & $88.2\pm1.3$ & $100.7\pm0.5$ \\
\hline \hline
Ni, $3\times3$ & $\infty$ & $1$ & $0$ & $0$ & $0$ & $34.9\pm0.4$ & $34.9\pm0.4$ \\
\hline
Ni, $3\times3$ & $1.27$ & $0.97\pm0.01$ & $0.12\pm0.01$ & $0.04\pm0.01$ & $0.02\pm0.01$ & $33.1\pm0.8$ & $32.4\pm0.3$ \\
\hline
Ni, $3\times3$ & $1.11$ & $0.94\pm0.02$ & $0.15\pm0.02$ & $0.07\pm0.02$ & $0.04\pm0.02$ & $29.5\pm1.3$ & $30.0\pm0.4$ \\
\hline
Ni, $3\times3$ & $0.95$ & $0.80\pm0.05$ & $0.14\pm0.05$ & $0.08\pm0.05$ & $0.06\pm0.05$ & $22.9\pm0.9$ & $20.4\pm0.7$ \\
\hline
Ni, $4\times4$ & $0.95$ & $0.82\pm0.02$ & $0.15\pm0.03$ & $0.08\pm0.03$ & $0.06\pm0.03$ & $19.9\pm1.3$ & $21.6\pm0.7$ \\
\hline
Ni, $4\times4$ & RMC & $0.98\pm0.01$ & $0.31\pm0.01$ & $0.12\pm0.01$ & $0.09\pm0.01$ & $25.8\pm0.8$ & $33.2\pm0.3$ \\
\hline
\end{tabular}
\end{table*}

As mentioned above, short-range order can sometimes have a significant effect
on resistivity. In this section we analyze the effect of MSRO on SDR in Fe and
Ni. In particular, it is important to check whether the large disagreement with
experiment for Ni found in Section \ref{MFA} can be due to the use of MFA
which neglects MSRO. This is especially interesting
because strong MSRO in Ni has been suggested by some experiments
\cite{Mook,Moriya} and theories. \cite{Antropov,Wang}

Spin disorder configurations with MSRO were generated using the Monte Carlo
(MC) method for the classical Heisenberg model with nearest-neighbor (NN)
exchange interaction on bcc and fcc lattices (for Fe and Ni, respectively).
These configurations were used to calculate SDR as described above, which can
then be compared with MFA results. As before, we usually used $4\times4$ and
$3\times3$ supercells for Fe and Ni, respectively, which was sufficient to
achieve Ohmic scaling of the conductance.

The results of SDR calculations for this model are shown in Fig.\ \ref{SDR}.
The MC results for SDR are very close to MFA results corresponding to the same
magnetization, in spite of the presence of MSRO in the MC model. We also
performed additional calculations with larger cell cross-sections ($6\times6$
for Fe and $4\times4$ for Ni) in order to check whether the finite-size effects
are important close to $T_c$. The results are shown by empty and full inverted
triangles in Fig.\ \ref{SDR}; the temperature was taken to be the same as for
the neighboring point for a smaller cross-section. It is clear that finite-size
effects have no appreciable effect on SDR. Thus, is appears that the spin-spin
correlations on the length scales comparable to our supercell lateral dimension
have negligible effect on SDR, in agreement with the conclusions of Ref.\
\onlinecite{Fisher}.

These results clearly show that MSRO characteristic for the NN Heisenberg model
has virtually no effect on SDR in Fe and Ni. The magnitude of MSRO is
illustrated in Table ~\ref{tab2} where the spin-spin correlators for the first
three shells of nearest neighbors are shown together with the corresponding
resistivities. While MSRO in the NN Heisenberg model for the close-packed
lattices considered here is not strong, it is seen that its effect on SDR is
much smaller even compared with the values of the NN spin-spin correlators.
This insensitivity is likely due to the averaging over all the electronic
states on the Fermi surface,\cite{Fisher} which should be very effective in
destroying the interference from scattering at different sites in transition
metals with complicated Fermi surfaces. In fact, this averaging is also
responsible for the small standard deviation of the local DOS from its mean
(Fig.\ \ref{Fe-dos}) and justifies the DLM approach for transition metals.

The spin-spin correlation function in real materials may be more complicated
than in the NN Heisenberg model. However, if the interaction has a longer range
while remaining mainly ferromagnetic, the MSRO must be \emph{weaker} compared
to the NN model.\cite{MFA} First-principles calculations for both ferromagnetic
and paramagnetic nickel show that the exchange parameters beyond nearest
neighbors, while being much smaller than the dominant NN exchange, stay mainly
ferromagnetic.\cite{AHS,Ruban} Interaction of this kind can not support
stronger MSRO compared to the NN Heisenberg model.

Since the complete insensitivity of the resistivity to MSRO within the NN
Heisenberg model is somewhat surprising, we have checked whether it is possible
to observe some change of SDR using spin configurations with an
artificially introduced stronger MSRO. For this purpose we used the reverse
Monte Carlo (RMC) method \cite{RMC} to produce a set of spin configurations
with almost zero magnetization and deliberately targeting strong MSRO in the NN
shell. Due to geometrical constraints, the spin-spin correlators in different
neighbor shells are not independent. We found it quite difficult to produce
strongly correlated nearest neighbors and at the same time avoid unphysical
artefacts in the long-range behavior of the correlation function. The spin-spin
correlators for the first three shells of neighbors in our RMC model are listed
in Table ~\ref{tab2}. The corresponding values of SDR calculated for Fe and Ni
with this set of spin configurations are also listed in Table~\ref{tab2} and
shown by full and empty diamonds in Fig.\ \ref{SDR}. Here we used $4\times4$
supercells for both Fe and Ni and checked for finite-size effects using
$6\times6$ supercells for Fe (essentially no difference was observed compared
to $4\times4$ cells). As seen in Table ~\ref{tab2}, the MSRO in this model is
significantly stronger compared to the NN Heisenberg model. The effect of this
strong MSRO on the calculated SDR is now noticeable but still relatively small;
the SDR is reduced compared to its MFA values by 12\% for Fe and 22\% for Ni.

\section{Effect of the local moment reduction}\label{reduc}

Reduction of the local moment is a universal feature of itinerant magnets as
revealed by spin fluctuations theories. \cite{Moriya} As discussed in Section
\ref{MFA}, the local moment in Fe is very stable and changes only slightly in
the paramagnetic state compared to zero temperature. Therefore, our
calculations based on the ground-state value of the local moment agree well
with experiment for Fe. Still, the SDR is sensitive to the local moment, and a
small reduction of it noticeably improved the agreement with experiment at
higher temperatures. Since the SDR was found to be insensitive to MSRO, it is
reasonable to attribute the large overestimation of the high-temperature SDR in
Ni to the neglect of the local moment reduction. Here we study this issue in
detail.

In the paramagnetic DLM state the local moment in Ni vanishes, \cite{Staunton}
but it is partially restored by longitudinal spin fluctuations.
\cite{Moriya,Ruban} Following the idea of separation of low and high-energy
fluctuations, we assume that the current-carrying quasiparticles near the Fermi
level experience the averaged exchange-correlation field generated by fast
longitudinal spin fluctuations, and that this ``mean field'' is adequately
represented by noncollinear DFT with disordered local moments constrained to
their square-averaged values. The atomic potentials are therefore obtained
using the fixed spin method \cite{fsmom} with the value of the constrained
local moment treated as an adjustable parameter, which has a physical meaning
and can be measured experimentally. Other approximations are, in principle,
possible; for example, the longitudinal spin fluctuations can be explicitly
included in the same noncollinear DFT approach, i.\ e.\ they can be considered
to be ``slow'' rather than ``fast.'' Since the separation in slow and fast
degrees of freedom is not well defined, we did not attempt to study the role of
these additional fluctuations.

The calculated paramagnetic SDR of Ni as a function of the local moment is
shown in Table ~\ref{tab3}. As seen, SDR is very sensitive to the value of the
local moment. Comparison with experimental SDR shows that our predicted value
of the square-averaged local moment in paramagnetic state of Ni is equal to
$0.35\mu_B$ (using the more accurate basis set with $l_{max}=3$).
Unfortunately, we are not aware of experimental measurements suitable for
comparison with this prediction.

\begin{table}
\caption{Spin-disorder resistivity in $\mu\Omega\cdot$cm for paramagnetic Ni as
a function of the fixed local moment. $2\times2$ supercells and basis sets with
$l_{max}=2$ and $l_{max}=3$ were used. Standard deviations of SDR due to
limited disorder sampling are included.} \label{tab3}
\begin{ruledtabular}
\begin{tabular}{lcccc|c}
Local moment, $\mu_B$ & $0.66$ & $0.5$ & $0.4$ & $0.3$ & Exp.\cite{Weiss} \\
\hline
$l_{max}=2$ & $34\pm0.6$ & $27\pm0.5$ & $21\pm0.4$ & & $15$ \\
$l_{max}=3$ & $29\pm0.6$\footnote{This value corresponds to unconstrained local moment of $0.63 \mu_B$.} & $23\pm0.5$ & $18\pm0.4$ & $12\pm0.3$ & $15$ \\
\end{tabular}
\end{ruledtabular}
\end{table}

\section{Discussion and conclusions}

Numerous previous studies \cite{Kasuya,dGF,Mannari,VI,Fisher} based on the
$s$-$d$ model concluded that SDR in the paramagnetic state is essentially
proportional to $J_{sd}^2S(S+1)$ where $S$ is the spin of the partially filled
$3d$ shell. This dependence is easy to understand based on the Fermi golden
rule with averaging over the initial states of the $3d$ spin. In our treatment
based on noncollinear DFT, the exchange-correlation field with randomized
directions on different sites plays the role of the $s$-$d$ Hamiltonian.
However, contrary to the $s$-$d$ model, the $3d$ spin is treated classically,
i.\ e.\ $\mathbf{S}$ is just a classical vector and not an operator. The Fermi
golden rule in our case would give a paramagnetic SDR proportional to
$J_{sd}^2S^2$. Thus, if the $S(S+1)$ factor were correct, noncollinear DFT
calculations would underestimate the paramagnetic SDR by a factor $(S+1)/S$.
This factor is close to 2 for Fe and more than 3 for Ni. In reality, the
calculated SDR agrees well with experiment for Fe and is \emph{over}estimated
for Ni (if the local moment reduction is not included). We believe that these
results provide clear evidence against the $S(S+1)$ factor which appears if the
local moments are treated as local atomic spins. Instead, the classical
description of the local magnetic fluctuations in the spirit of the DLM
approach is supported by our results. We suggest that the itinerancy of the
$3d$ electrons is crucial for this behavior. Qualitatively, one can argue that
the low-energy fluctuations in Fe or Ni on the scale of $kT$ (which the
resistivity is most sensitive to) are similar to classical rotations of the
local moments rather than quantum fluctuations of localized spins. It would be
interesting to investigate this issue for magnets with a varying degree of
localization, including rare-earth systems.

Some poorly controlled assumptions are involved in the extraction of $\rhom$
from the experimental data.\cite{Weiss} First, it is assumed that $\rhom$ is
constant in the wide temperature range above $T_c$ where the total resistivity
is linear in $T$. This assumption implies that the local moments (or at least
their mean-squared average) are constant in this range. Spin fluctuation
theories for itinerant metals show that the local moments may change with
temperature above $T_c$. \cite{Moriya,Ruban,WGB} Such change will contribute to
the slope of $\rho$ above $T_c$, and hence the separation of $\rhom$ from the
phonon contribution would be inaccurate.

On the other hand, it has been argued that the phonon contribution to the
resistivity may be sensitive to spin disorder, because the latter may change
the character of states at the Fermi level.\cite{Coles,Mott} In particular, in
Ni the filled majority-spin $d$ states may be lifted up to the Fermi level by
spin disorder, thereby facilitating interband $s$-$d$ scattering by phonons.
This effect may therefore introduce an unusual temperature dependence of the
phonon contribution, which makes spin disorder and phonon effects non-additive,
even if the scattering rates themselves obey Matthiessen's rule. Since we have
not studied this effect here, our comparison of SDR with experiment for Ni is
incomplete. However, the phonon contribution can be expected to follow the
Bloch-Gr\"uneisen form above $T_c$ with the electron-phonon scattering
renormalized by spin disorder; therefore, the influence of spin disorder on the
phonon contribution should not invalidate the procedure used for subtracting
this contribution above $T_c$.

In conclusion, we have calculated the spin-disorder resistivity of Fe and Ni in
the whole temperature range up to $T_c$ using both the mean-field approximation
and the nearest-neighbor Heisenberg model to represent the canonical ensemble
of classical spin configurations. We found that SDR is insensitive to the
magnetic short-range order (MSRO) in Fe and Ni. The SDR in Fe depends linearly
on $M^2(T)$ which implies that the main effect of spin disorder is to introduce
scattering, which is proportional to the variance of the random potential. For
Ni the calculated temperature dependence is more complicated; at elevated
temperatures close to $T_c$ the SDR grows faster than expected. This faster
increase of SDR may be explained by the reduction of the exchange splitting
which lifts the heavy bands up to the Fermi level, thereby increasing the
scattering rate. The results for Fe are in very good agreement with experiment
if the atomic potentials are taken from zero temperature and frozen, but for Ni
the SDR calculated in this way is strongly overestimated. This disagreement is
attributed to the reduction of the local magnetic moment in Ni. Comparison with
experimental SDR leads to a value of $0.35 \mu_B$ above $T_c$, which may be
compared with experiment.

\acknowledgments

We are grateful to V.\ P.\ Antropov and E.\ Y.\ Tsymbal for useful discussions.
This work was supported by the Nebraska Research Initiative, NSF EPSCoR, and
NSF MRSEC. K.\ D.\ B.\ is a Cottrell Scholar of Research Corporation. The work
was completed utilizing the Research Computing Facility of the University of
Nebraska-Lincoln. A portion of this research at Oak Ridge National Laboratory's
Center for Nanophase Materials Sciences was sponsored by the Scientific User
Facilities Division, Office of Basic Energy Sciences, U.S. Department of
Energy.


\begin{thebibliography}{9}
\bibitem{MFA} The MFA criterion of validity is $1/z\ll1$ where $z$ is the
effective number of neighbors inside the interaction range.
\end{thebibliography}
\end{document}